\documentclass{sigchi-ext}
\usepackage[T1]{fontenc}
\usepackage{textcomp}
\usepackage[scaled=.92]{helvet} 
\usepackage{graphicx} 
\usepackage{balance}  
\usepackage{booktabs} 
\usepackage{ccicons}  
\usepackage{ragged2e} 

\usepackage[all]{nowidow}

\usepackage[utf8]{inputenc} 

\def\plaintitle{Scientific Outreach with Teegi, a Tangible EEG Interface to Talk about Neurotechnologies} \def\plainauthor{Jérémy Frey, Renaud Gervais, Thibault Lainé, Maxime Duluc, Hugo Germain, Stéphanie Fleck, Fabien Lotte, Martin Hachet}
\def\plainkeywords{Tangible Interaction; EEG; BCI; Scientific Outreach}

\title{Scientific Outreach with Teegi, a Tangible EEG Interface to Talk about Neurotechnologies}

\numberofauthors{8}
\author{%
  \alignauthor{%
    \textbf{Jérémy Frey}\\
    \affaddr{Inria, France}\\
    \affaddr{Ullo, France}\\
    \email{jeremy.frey@inria.fr} } \alignauthor{%
    \textbf{Renaud Gervais}\\
    \affaddr{Inria, France}\\
    \email{renaud.gervais@inria.fr} } \vfil \alignauthor{%
    \textbf{Thibault Lainé}\\    
    \affaddr{Inria, France}\\
    \email{thibault.laine@inria.fr} } \alignauthor{%
    \textbf{Maxime Duluc}\\
    \affaddr{Inria, France}\\
    \email{maxime.duluc@inria.fr} } \vfil \alignauthor{%
    \textbf{Hugo Germain}\\
    \affaddr{Inria, France}\\
    \email{hugo.germain@inria.fr} } \alignauthor{%
    \textbf{Stéphanie Fleck}\\
    \affaddr{Univ. Lorraine, France}\\
    \email{stephanie.fleck@univ-lorraine.fr} } \vfil \alignauthor{%
    \textbf{Fabien Lotte}\\
    \affaddr{Inria, France}\\
    \email{fabien.lotte@inria.fr} } \alignauthor{%
    \textbf{Martin Hachet}\\
    \affaddr{Inria, France} \\
    \affaddr{martin.hachet@inria.fr} }
 }

\definecolor{linkColor}{RGB}{6,125,233}
\hypersetup{%
  pdftitle={\plaintitle},
  pdfauthor={\plainauthor},
  pdfkeywords={\plainkeywords},
  bookmarksnumbered,
  pdfstartview={FitH},
  colorlinks,
  citecolor=black,
  filecolor=black,
  linkcolor=black,
  urlcolor=linkColor,
  breaklinks=true,
}

\copyrightinfo{Permission to make digital or hard copies of part or all of this work for personal or classroom use is granted without fee provided that copies are not made or distributed for profit or commercial advantage and that copies bear this notice and the full citation on the first page. Copyrights for third-party components of this work must be honored. For all other uses, contact the Owner/Author.\\
Copyright is held by the owner/author(s).\\
{\emph{CHI'17 Extended Abstracts}}, May 06-11, 2017, Denver, CO, USA.\\
ACM 978-1-4503-4656-6/17/05. \\
http://dx.doi.org/10.1145/3027063.3052971}

\begin{document}

\maketitle

\RaggedRight{} 


\begin{abstract}

Teegi is an anthropomorphic and tangible avatar exposing a users' brain activity in real time. It is connected to a device sensing the brain by means of electroencephalography (EEG). Teegi moves its hands and feet and closes its eyes along with the person being monitored. It also displays on its scalp the associated EEG signals, thanks to a semi-spherical display made of LEDs. Attendees can interact directly with Teegi -- e.g. move its limbs -- to discover by themselves the underlying brain processes. Teegi can be used for scientific outreach to introduce neurotechnologies in general and brain-computer interfaces (BCI) in particular.

\end{abstract}


\category{I.2.6 }{Learning}{Knowledge acquisition}

\begin{figure}[!ht]
  \includegraphics[width=0.9\columnwidth]{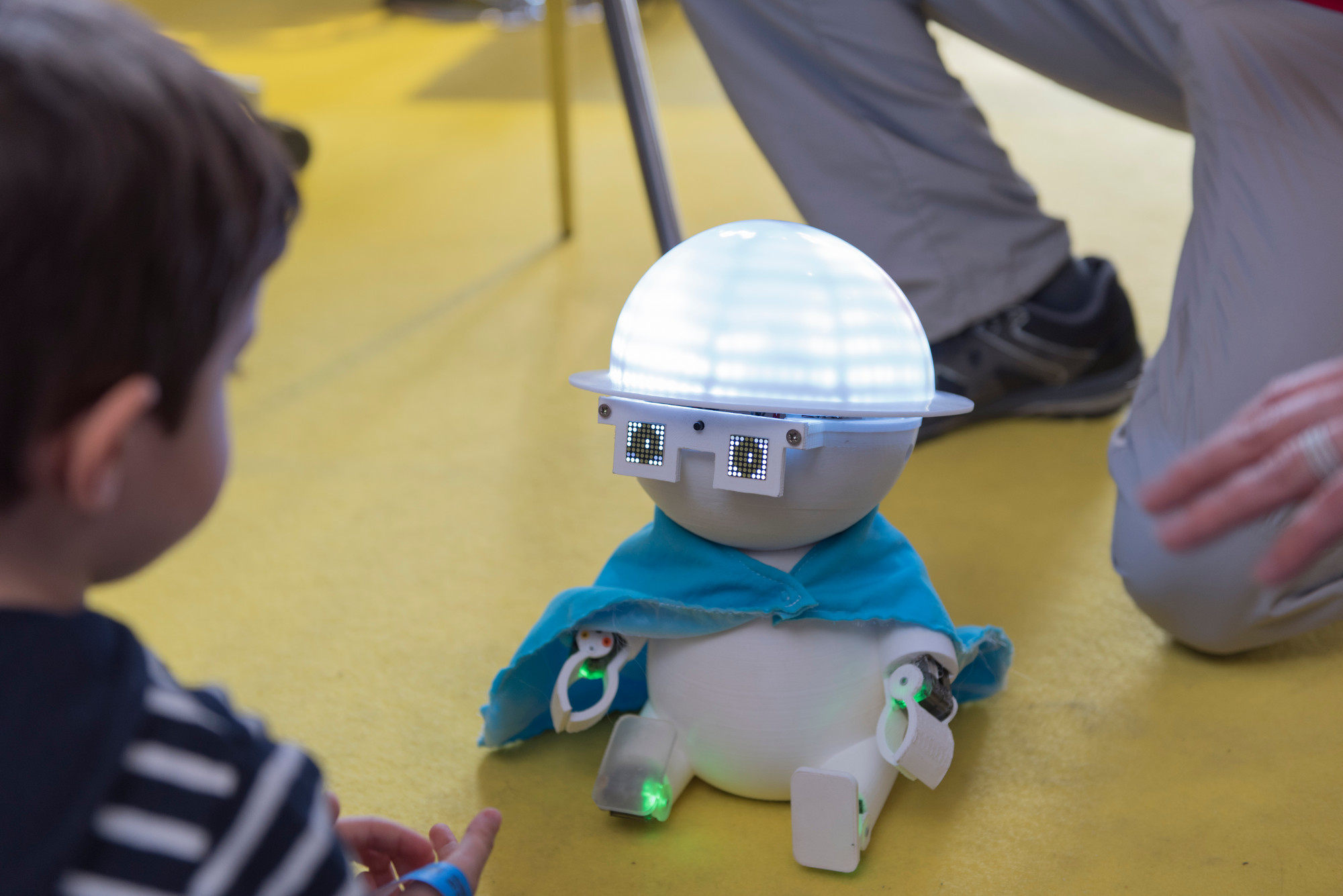}
  \caption{Teegi displays brain activity in real time by means of electroencephalography. It can be used to explain to novices or to children how the brain works.}~\label{fig:teegi-close}
\end{figure}

\section{Introduction}

\begin{figure}[!ht]
  \includegraphics[width=0.9\columnwidth]{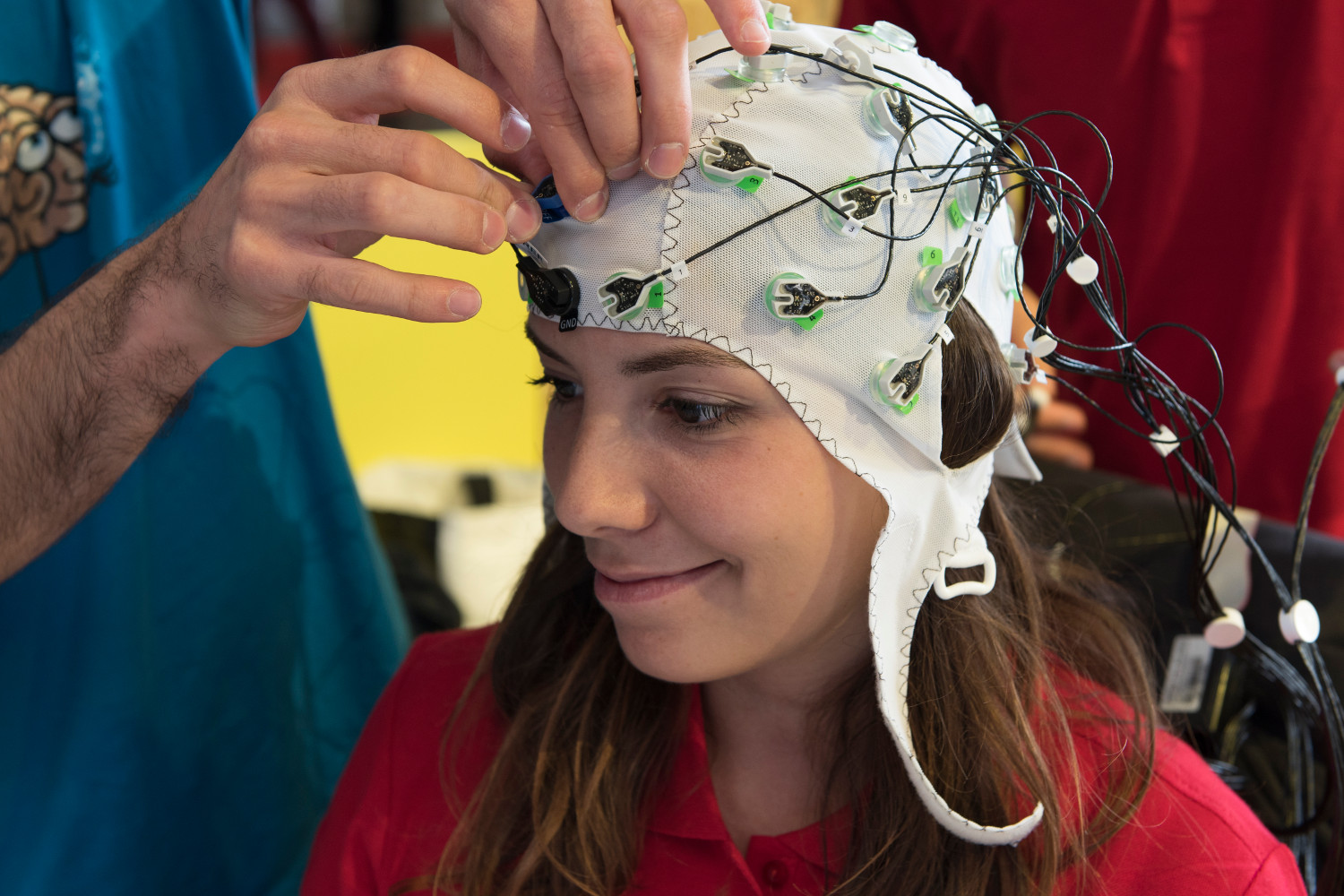}
  \caption{An electroencephalography (EEG) cap.}~\label{fig:eeg}
\end{figure}

Teegi (Figure \ref{fig:teegi-close}) is a Tangible ElectroEncephaloGraphy (EEG) Interface that enables novice users to get to know more about something as complex as neuronal activity, in an easy, engaging and informative way. Indeed, EEG measures the brain activity under the form of electrical currents, through a set of electrodes placed on the scalp and connected to an amplifier (Figure \ref{fig:eeg}). EEG is widely used in medicine for diagnostic purposes and is also increasingly explored in the field of Brain-Computer Interfaces (BCI). BCIs enable a user to send input commands to interactive systems without any physical motor activities or to monitor brain states \cite{Pfurtscheller2001,Frey2016}. For instance, a BCI can enable a user to move a cursor to the left or right of a computer screen by imagining left or right hand movements respectively.

BCI is an emerging research area in Human-Computer Interaction (HCI) that offers new opportunities. Yet, these emerging technologies feed into fears and dreams in the general public (``telepathy'', ``telekinesis'', ``mind-control'', ...). Many fantasies are linked to a misunderstanding of the strengths and weaknesses of such new technologies. Moreover, BCI design is highly multidisciplinary, involving computer science, signal processing, cognitive neuroscience and psychology, among others. As such, fully understanding and using BCI can be difficult. 
 
In order to mitigate the misconceptions surrounding EEG and BCI, we introduced Teegi in \cite{Frey2014b}, as a new system based on a unique combination of spatial augmented reality, tangible interaction and real-time neurotechnologies. With Teegi, a user can visualize and analyze his or her own brain activity in real-time, on a tangible character that can be easily manipulated, and with which it is possible to interact.

\begin{figure}[!ht]
  \includegraphics[width=0.5\columnwidth]{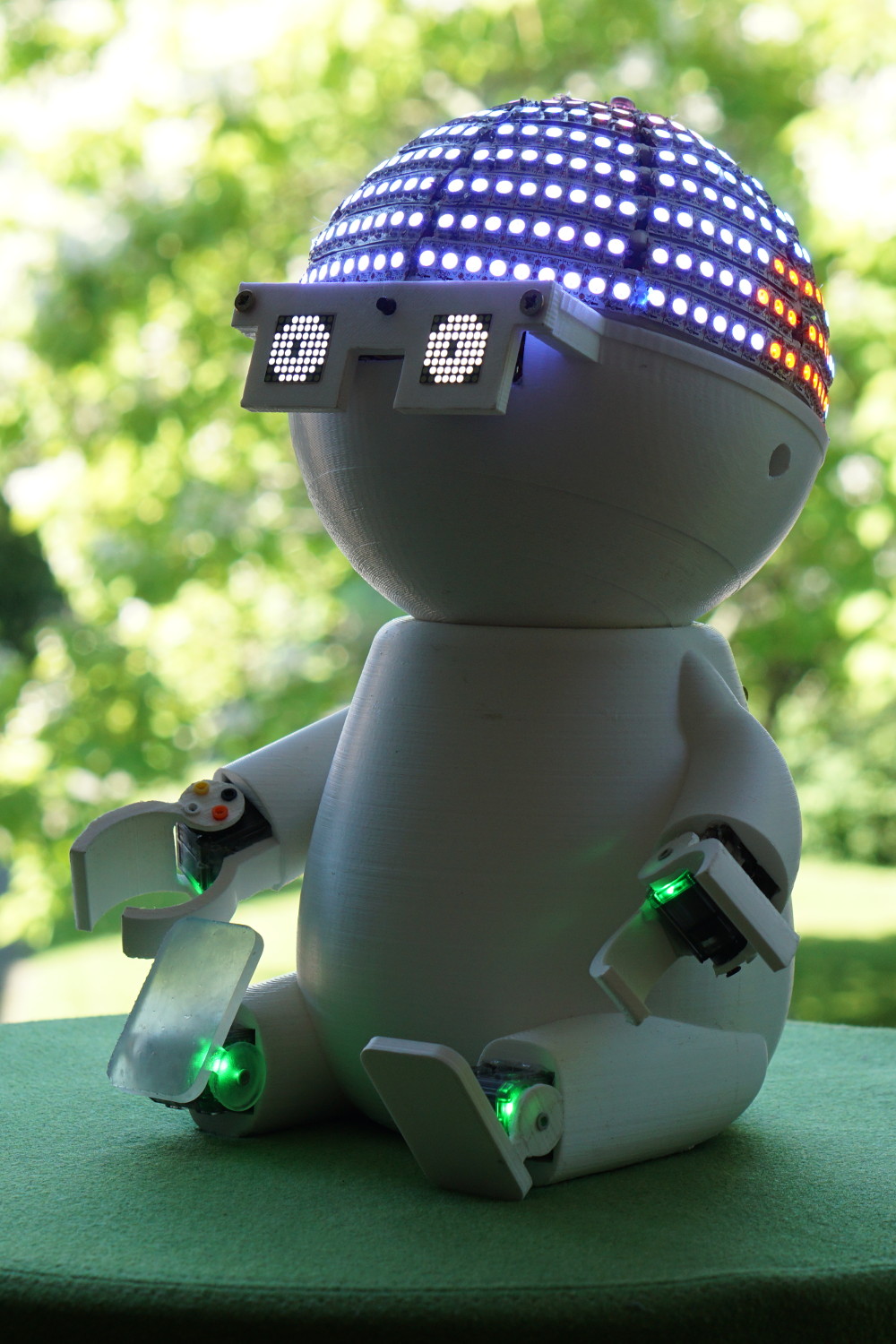}
  \includegraphics[width=0.5\columnwidth]{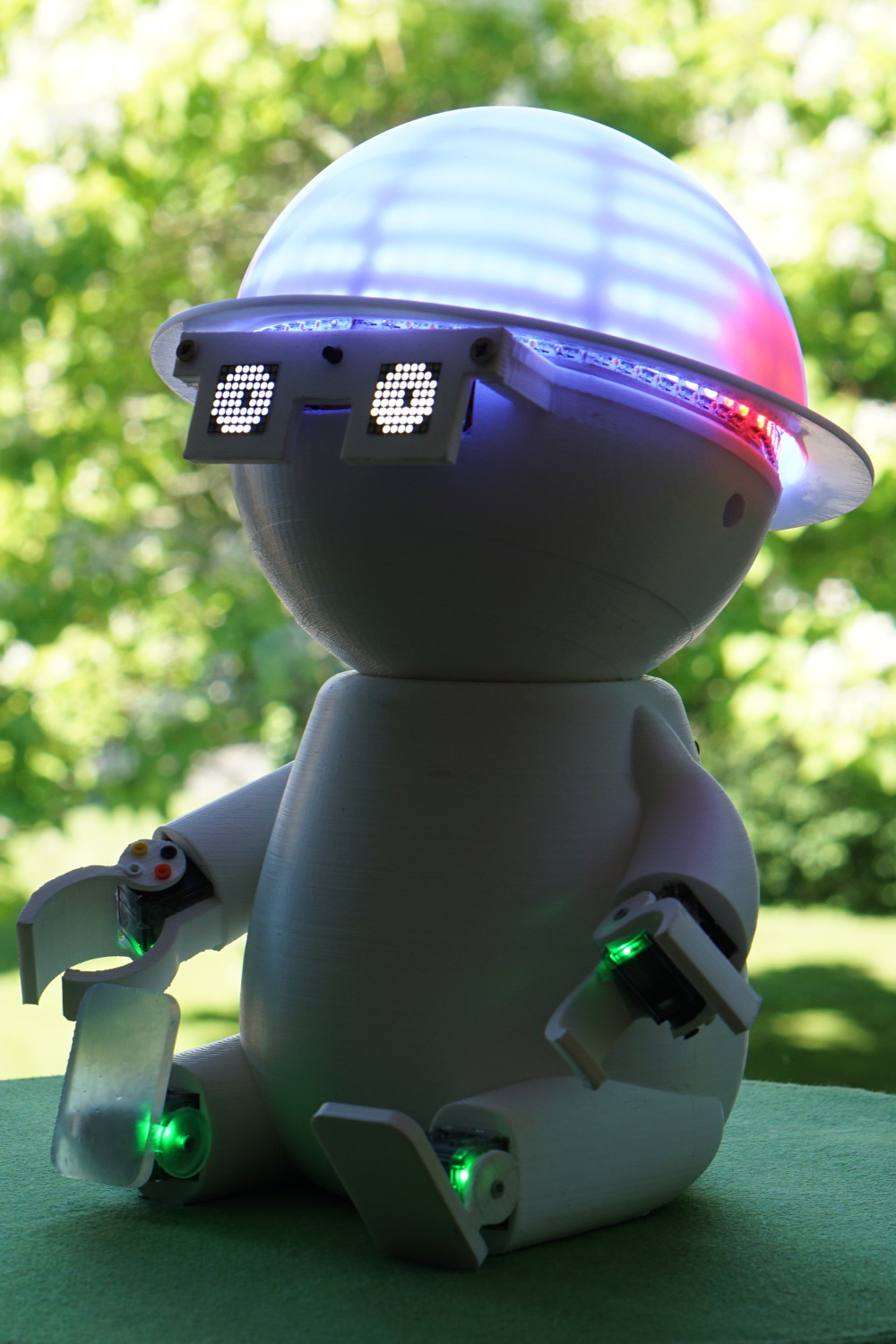}
  \caption{Teegi possesses a semi-spherical display composed of 402 LEDs (left) which is covered by a layer of acrylic glass (right).}~\label{fig:leds}
\end{figure}

Since this first design, we switched from projection-based and 3D tracking technologies to a LEDs-based semi-spherical display (Figure \ref{fig:leds}). All the electronics are now embedded. This way, Teegi became self-contained and can be easily deployed outside the lab. We also added servomotors to Teegi, so that he can move and be moved. This way, we can more intuitively describe how hands and feet movements are linked to specific brain areas and EEG patterns. 

Our first exploratory studies in the lab shown that interacting with Teegi seemed to be easy, motivating, reliable and informative. Since then, we confirmed that Teegi is a relevant training and scientific outreach tool for the general public. Teegi as a ``puppet'' -- an anthropomorphic augmented avatar -- proved to be a good solution in the field to break the ice with the public and explain complex phenomena to people from all horizons, from children to educated adults. We tested Teegi across continents and cultures during scientific fairs before thousands of attendees, in India as well as in France. 

\section{Description of the system}

The installation is composed of three elements: the EEG system that records brain signals from the scalp, a computer that processes those signals and the puppet Teegi, with which attendees interact.

EEG signals can be acquired from various amplifiers, from medical grade equipment to off-the-shelf devices. The choice of system mainly depends on the brain states that one wants to describe through Teegi. For instance, our installation focuses on the brain areas involved in motor activity, hence we require electrodes over the parietal zone. We use Brain Products' LiveAmp\footnote{\url{http://www.brainproducts.com/}} and Neuroelectrics' Enobio\footnote{\url{http://www.neuroelectrics.com/}} systems. The former has 32 gel-based electrodes, which give more accurate readings but are more tedious to setup. The Enobio has 20 ``dry'' electrodes, making it easier to switch the person whose brain activity is being monitored, but it is more prone to artifacts -- e.g. if the person is not sitting. Both those systems are mobile and wireless.

The readings are sent to a computer. Those signals are acquired and processed by OpenViBE\footnote{\url{http://openvibe.inria.fr/}}, an open-source software dedicated to BCIs. OpenViBE acts as an abstraction layer between the amplifiers and Teegi, sending processed EEG signals through wifi to Teegi -- for more technical details about the signal processing, see \cite{Frey2014b}.

Teegi is 3D printed, designed to be both attractive and hold the various electronic components. It embeds a Raspberry Pi 3 and NiMh batteries (autonomy of approximately 2 hours). A python script on the Raspberry Pi handles the 402 LEDs (Adafruit Neopixel) covering the ``head'', which are connected to its GPIO pins. For a smoother display, the light of the LEDs is diffused by a 3mm thick cap made of acrylic glass. Two 8-by-8 white LEDs matrices picture the eyes. The script also commands the servomotors placed in the hands and feet, 4 Dynamixel XL320. 

\section{Scenario}

Teegi possesses two operating modes: avatar and puppet. As an avatar, it uses the EEG system and directly translates the brain states being recorded into movements and brain activity display. As a puppet,  the EEG is not used and one could interact freely with Teegi (move its limbs, close its eyes with a trigger), as a way to discover which brain regions are involved in specific motor activities or in vision.

Typically, a demonstration of Teegi starts by letting the audience play with the puppet mode. When one closes Teegi's eyes, she would notice that the display changed in the ``back'' of the head. We then explain that the occipital area holds the primary visual cortex. When ones move the left hand, a region situated on the right part of Teegi's scalp is illuminated. When the right hand is moved it is the opposite, LEDs situated on the left turn blue or red. We take this opportunity to explain that the body is contralaterally controlled; the right hemisphere controls the left part of the body and vice versa. Depending on the nature of the attendees, we can go further and explain the phenomenon of desynchronization that takes place within the motor cortex when there is a movement, and the synchronization that occurs between neurons when it ends.  

With few intuitive interactions, Teegi is a good mediator for explaining basic neuroscience. When used as an avatar, the LED display and Teegi's servomotors are linked to the EEG system -- for practical reasons one of the demonstrators wear the EEG cap. We demonstrate that when the EEG user closes her eyes, Teegi closes his. Moreover, Teegi's hands and feet move according to the corresponding motor activity (real or imaged) detected in the EEG signal. During the whole activity, Teegi's brain areas are illuminated according to the real-time EEG readings.

\section{Audience and Relevance}

The demonstration is suitable for any audience: students, researchers, naive or expert in BCI. We would like to meet with our HCI pairs to discuss the utility of tangible avatars that are linked to one's physiology. We believe that such interfaces, promoting self-investigation and anchored in reality, are a good example of how the field could contribute to education (e.g. \cite{Horn2009}) -- moreover when it comes to rather abstract information. Teegi could also foster discussions about the pitfalls of BCI; for example it is difficult to avoid artifacts and perform accurate brain measures. 

Overall, Teegi aims at deciphering complex phenomenon as well as raising awareness about neurotechnologies. Beside scientific outreach, in the future we will explore how Teegi could be used to better learn BCIs and, in medical settings, how it could help to facilitate stroke rehabilitation.

\section{Acknowledgments}

We want to thank Jérémy Laviole and Jelena Mladenović for their help and support during this project.

\balance{} 

\bibliographystyle{SIGCHI-Reference-Format}
\bibliography{biblio}

\end{document}